\documentclass[aps,twocolumn,groupaddress,usenatbib,nofootinbib,showkeys,showpacs,altaffilletter]{revtex4}

\usepackage{graphicx}
\usepackage{dcolumn}
\usepackage{bm}
\usepackage{latexsym}
\usepackage{mathrsfs}

\begin{document}

\title{A tensor instability in the Eddington inspired Born-Infeld Theory of Gravity}

\author{Celia Escamilla-Rivera}
\email{celia\_escamilla@ehu.es}
\affiliation{Fisika Teorikoaren eta Zientziaren Historia Saila, Zientzia
eta Teknologia Fakultatea, Euskal Herriko Unibertsitatea, 644 Posta
Kutxatila, 48080, Bilbao, Spain.}
\affiliation{Astrophysics, University of Oxford, DWB, Keble Road, Oxford, OX1 3RH, UK}

\author{M\'aximo Banados}
 \email{maxbanados@fis.puc.cl}
 \affiliation{P. Universidad Cat\'olica de Chile, Avenida Vicuna Mackema 4860,
 Santiago, Chile}

\author{Pedro G. Ferreira}
\email{p.ferreira1@physics.ox.ac.uk}
\affiliation{Astrophysics, University of Oxford, DWB, Keble Road, Oxford, OX1 3RH, UK}


\begin{abstract}

In this paper we consider an extension to Eddington's proposal for the gravitational action. We study tensor perturbations of a homogeneous and isotropic space-time in the Eddington regime, where modifications to Einstein gravity are strong. We find that the
tensor mode is linearly unstable deep in the Eddington regime and discuss its cosmological
implications.

\end{abstract}

\keywords{Cosmology, Gravitational Waves, Graviton.}

\pacs{$98.80.-k,04.30.-w,14.70.Kv$}


\maketitle

\section{Introduction}

An alternative theory of gravity was recently proposed in \cite{Banados:2010ix} which
attempted to extend Eddington's affine theory of gravity to the matter sector \cite{Eddington1923} (see also \cite{Deser:1998rj,Vollick04}).
The new theory is formulated in Palatini form in terms of the affine connection $\Gamma^{\mu}_{\phantom{\mu}\alpha\beta}$ and a space-time
metric, $g_{\alpha\beta}$ such that the gravitational action is given by
\begin{eqnarray}\label{palatini}
S_{EBI}[g,\Gamma,\Psi]&=& \frac{2}{\kappa}\int{d^4 x \left[\sqrt{\left|g_{\mu\nu}
+\kappa R_{\mu\nu}(\Gamma)\right|} -\lambda\sqrt{g}\right]} \nonumber \\
&& + S_{m}[g,\Psi], \nonumber
\end{eqnarray}
where $\kappa =8\pi G$, $\Psi$
denotes any additional matter fields. $R_{\mu\nu}$ is the symmetric Ricci
tensor constructed with $\Gamma$. This action can reproduce  Eddington's original action at large values of $\kappa R$ and Einstein's at small values.

It was shown in \cite{Banados:2010ix} that this theory has a novel behaviour within sources.
It was found that, in stars, compact objects and black holes, that the Eddington regime might
not only lead to the avoidance of singularities but to significant modifications to the
standard results of stellar astrophysics. The authors of \cite{Casanellas2012} showed that it was possible to test the Eddington corrections to Newtonian gravity using Solar physics, while in \cite{Pani:2011mg} it was shown that it should be possible to do the same around compact rotating sources.
In \cite{Avelino2012} it was shown that the mere existence of bound gravitational objects of a certain size led to stringent constraints on the free parameter of the theory while in \cite{Delsate2012} it was shown that the
theory could re-expressed as a bigravity theory, along the lines of \cite{BGRS2009}.

The Eddington regime will also arise in the very early Universe and it was shown in
\cite{Banados:2010ix} that it may have led to a minimum scale factor. Hence, and more significantly than in black holes, the Eddington regime seems to prevent the formation of
cosmological singularities, at least when seen in the context of homogeneous and isotropic space times.

In this brief note, we look at one particular aspect of the cosmological behaviour of the Eddington regime by studying the structure and evolution of linear tensor mode perturbations,  i.e. cosmological gravitational waves. We focus on tensor modes because they are purely gravitational, unlike the scalar modes that are a combination of metric and density perturbations. This note is structured as follows: in section \ref{theory} we lay out the main equations and derive the evolution equation for traceless, transverse perturbations (i.e. tensor modes) to
a homogeneous and isotropic space time; in section \ref{evolution} we find the solutions to the tensor equations in the Eddington regime and show that there is an instability; in section \ref{conclusions} we discuss the implication of our findings for this specific model and the wider consequences of this tensor
instability for bigravity theories and their equivalents.

\section{Tensor modes in the Eddington Born Infeld Theory}
\label{theory}
In this section we review some key features of \cite{Banados:2010ix} and
introduce the procedure that will lead us to compute the tensor
perturbations using the auxiliary metric $q_{\mu\nu}$.
In Eq.(\ref{palatini}) we see that the metric and the connection are
treated as independent variables, known as Palatini
variation. The resulting new set of evolution equations can be conveniently written in the form \cite{Banados:2010ix}
\begin{eqnarray}\label{evolution1}
\sqrt{\left|\frac{q}{g}\right|}(q^{-1})^{\mu\nu} - \lambda g^{\mu\nu}&=&-\kappa T^{\mu\nu}(g,\Psi),\\
\label{evolution3}
q_{\mu\nu}-  g_{\mu\nu}&=&\kappa R_{\mu\nu}(q),
\end{eqnarray}
where $\lambda=1+\kappa\Lambda$, and $\kappa$ is a constant with the inverse
dimensions to that of the cosmological constant $\Lambda$. Here $q_{\mu\nu}$ is an auxiliary rank-2 tensor, related to the original connection $\Gamma$ via the Christoffel symbol.  $R_{\mu\nu}(q)$ is the associated curvature. The energy-momentum tensor is coupled to the metric $g_{\mu\nu}$. These equations imply that $T^{\mu\nu}$ is conserved in the usual sense, $T^{\mu\nu}_{ \ \ \ ; \nu}=0$, where the covariant derivative is taken with respect to the metric $g_{\mu\nu}$. This property is not obvious from the equations of motion, but follows in a straightforward way from the Lagrangian using invariance under general coordinate transformations.

We can consider a perturbed homogeneous and isotropic space time by choosing the two metrics to be of the form
\begin{eqnarray}\label{eq:metrics1}
g_{\mu\nu}dx^\mu dx^\nu=-a^2 d\eta^2 +a^2\left(\delta_{ij} +h_{ij}\right) dx^i dx^j ,
\end{eqnarray}
\begin{eqnarray}\label{eq:metrics2}
q_{\mu\nu}dx^\mu dx^\nu=-X^2 d\eta^2 +Y^2 \left(\delta_{ij}+\gamma_{ij}\right)dx^i dx^j ,
\end{eqnarray}
where $a$, $X$ and $Y$ are solely function of conformal time, $\eta$, both $h_{ij}$ and $\gamma_{ij}$ are transverse and traceless, i.e. $h_{ii}=\gamma_{ii}=0$ and
${\partial}_{i}h^{ij}={\partial}_{i}\gamma^{ij}=0$.

We can take the energy momentum tensor to be given by
\begin{equation}\label{energy-moment}
T_{\mu\nu}=(\rho+P) u_{\mu}u_{\nu}+Pg_{\mu\nu},
\end{equation}
where  $u^\mu=(1,0,0,0)$.
The system (\ref{evolution1})-(\ref{evolution3}) can then be used to find the evolution equations at all times. We will be focusing on the very early Universe, when matter could be described in terms of
a relativistic perfect fluid. It was shown in \cite{Banados:2010ix} that the modified Friedman-Robertson-Walker equation has the form:
\begin{eqnarray}
 3H^2 &=&\frac{1}{\kappa}\left[\kappa\rho -1 +\frac{1}{3\sqrt{3}}\sqrt{(\kappa\rho +1)(3-\kappa\rho)^3}\right] \nonumber \\
&& \times \left[\frac{(1+\kappa\rho)(3-\kappa\rho)^2}{(3+\kappa^2 \rho^2)^2}\right],
\end{eqnarray}
where $H=a'/a^2$ and for $\kappa\rho <<1$ we have $H^2 \simeq \rho/3$. This equation has critical points
for $H(\rho_{B})=0$ at a maximum density $\rho_{B} $ which depends on the sign of $\kappa$.
It is useful to write out the background equation that arise from the field equations:
\begin{eqnarray}
-\frac{|XY^3|}{X^2a^4}+\frac{\lambda}{a^2}&=&-\kappa\frac{\rho}{a^2}, \nonumber \\
\frac{|XY^3|}{Y^2a^4}-\frac{\lambda}{a^2}&=&-\kappa\frac{P}{a^2}. \label{back2}
\end{eqnarray}

To construct the perturbed field equations we need the following identities:
\begin{eqnarray}
\sqrt{|q|}&=&|XY^3|(1+{\cal O}(\gamma^2)), \nonumber \\
\sqrt{|g|}&=&|XY^3|(1+{\cal O}(h^2)), \nonumber \\
(q^{-1})^{ij}&=&\frac{1}{Y^2}(\delta^{ij}-\gamma^{ij}) \ \ \mbox{where} \quad \gamma^{ij}=\delta^{ik}\delta^{jl}\gamma_{kl}, \nonumber \\
(g)^{ij}&=&\frac{1}{a^2}(\delta^{ij}-h^{ij}) \ \ \mbox{where} \quad h^{ij}=\delta^{ik}\delta^{jl}h_{kl}, \nonumber \\
\delta T^{ij}&=& -\frac{P}{a^2}h^{ij}. \nonumber
\end{eqnarray}
We obtain the field equations in the form:
\begin{eqnarray}
-\frac{XY^3}{a^4}\frac{1}{Y^2}\gamma^{ij}+\frac{\lambda}{a^2}h^{ij}=\kappa\frac{P}{a^2}h^{ij}. \nonumber
\end{eqnarray}
But using Eq.(\ref{back2}) we have that
\begin{eqnarray}
-\frac{XY^3}{a^4}\frac{1}{Y^2}\gamma^{ij}+\frac{\lambda}{a^2}h^{ij}=\left(-\frac{XY^3}{a^4Y^2}+\frac{\lambda}{a^2}\right) h^{ij}, \nonumber
\end{eqnarray}
which simplifies greatly to
\begin{eqnarray}
\gamma_{ij}=h_{ij}. \nonumber
\end{eqnarray}
This is an intriguing result. Even though the tensor perturbations in both the metric and the auxiliary metrics are multiplied by different conformal factors, they are identical in this theory. Furthermore, even in the Einstein regime, where $X=Y=a$, we find that $\gamma_{ij}$ is non trivial and completely locked to the behaviour of $h_{ij}$.

We can now proceed to construct the evolution equation for $h_{ij}$ to find
\begin{eqnarray}
\nonumber
&& h''_{ij} +\left(3\frac{Y'}{Y}-\frac{X'}{X}\right)h'_{ij}+ \left[4\left(\frac{Y'}{Y}\right)^2 +2\frac{Y''}{Y}
-2\frac{X'}{X}\frac{Y'}{Y}\right. \\
&& \left.
-\frac{2}{\kappa}\left(\frac{X^2 a^2}{Y^2}-X^2 \right) +\left(\frac{X}{Y}\right)^2 k^2 \right] h_{ij} =0. \label{pert0}
\end{eqnarray}
The spatial background field equation can now be used to find
\begin{eqnarray}
Y^2=a^2+\kappa\frac{Y^2}{X^2}\left[\frac{Y''}{Y}+2\left(\frac{Y'}{Y}\right)^2-\frac{Y'X'}{YX}\right], \nonumber
\end{eqnarray}
which when replaced in Eq.(\ref{pert0}) leads to our final expression:
\begin{eqnarray}
h_{ij}''+\left(3\frac{Y'}{Y}-\frac{X'}{X}\right)h_{ij}'+\left(\frac{X}{Y}\right)^2k^2h_{ij}=0. \label{gwave}
\end{eqnarray}
This is a remarkably simple evolution equation for the tensor mode. In the Einstein limit it reduces to
\begin{eqnarray}
h_{ij}''+2\frac{a'}{a}h_{ij}'+k^2h_{ij}=0, \label{Einsteingw}
\end{eqnarray}
as expected.
\section{Evolution of Tensor modes}
\label{evolution}
We now wish to see how this system evolves in the different regimes. In the Einstein regime we
find that the evolution is indistinguishable from Einstein gravity, even though the auxiliary metric
is perturbed and present. The solutions to Eq.(\ref{Einsteingw}) will be the standard ones,
depending on the dominant source of energy momentum and can be found in any standard text book.
For example in the radiation era, we find that
\begin{eqnarray}
h_{ij}\propto \frac{1}{\sqrt{\eta}}{\cal H}^{(1)}_{\frac{1}{2}}(k\eta),\frac{1}{\sqrt{\eta}}{\cal H}^{(2)}_{\frac{1}{2}}(k\eta), \nonumber
\end{eqnarray}
where ${\cal H}^{(m)}_\nu(x)$ are Hankel functions of the $m$th kind. In the radiation era we have
$\nu=1/2$ and the solutions reduce to a regular ($\sin(k\eta)/k\eta$) and irregular ($\cos(k\eta)/k\eta$) solution as $k\eta\rightarrow 0$. The familiar behaviour
of gravitational waves \cite{Liddle:2000cg,Lidsey:1995np} emerges for very large $k\eta$ where we find decaying oscillatory solutions propagating at the speed of light. Crucial to this behaviour is that there is no time dependent factor multiplying the Laplacian (or
Fourier space analogue, $-k^2$).

It is in the Eddington regime that we find novel behaviour and we will first focus on $\kappa>0$.
In \cite{Scargill2012} it was shown that
the evolution of the background in the Eddington regime can be approximated by:
 \begin{eqnarray}
\frac{a}{a_B}&=&1+\exp\left(\sqrt{\frac{8}{3\kappa}}(t-t_0)\right), \nonumber \\
V\equiv\left(\frac{Y}{a}\right)^2&=&\sqrt{2}\exp\left(\frac{1}{2}(\sqrt{\frac{8}{3\kappa}}(t-t_0)\right), \nonumber \\
U\equiv \left(\frac{X}{a}\right)^2&=&\frac{1}{2}V^3 , \label{qfrw}
\end{eqnarray}
where $a_B$ is the minimum scale factor and $t$ is physical time. This primordial, non-singular behaviour was already alluded to in \cite{Banados:2010ix} and seems like an attractive, alternative explanation for early time.
It resurrects another of Eddington's ideas of doing away with  beginning and instead have the Universe indefinitely loitering in stasis in the distant past.
We can see, however, from Eq.(\ref{qfrw}) that the auxiliary metric {\it does} become singular as $t\rightarrow -\infty$ and it is this behaviour which will play a crucial role in the evolution of the tensor modes.

To proceed we need to re-express our background quantities in terms of conformal time, $\eta$ which we can find by integrating $d\eta=dt/a(t)$ using the above expression to find:
\begin{eqnarray}
{ a}=\frac{a_B}{1-\exp(\alpha\Delta\eta)}, \nonumber
\end{eqnarray}
where $\alpha=a_B\sqrt{8/(3\kappa)}$ and $\Delta\eta=\eta-\eta_i$ (where $\eta_i$ is a fiducial time) is negative (akin to when we write deSitter metric conformal coordinates we find that $\eta<0$). We then have that
\begin{eqnarray}
\frac{X^2}{Y^2}&=&\frac{\exp(\alpha\Delta\eta)}{1-\exp(\alpha\Delta\eta)}, \nonumber \\
\left(3\frac{Y'}{Y}-\frac{X'}{X}\right) &=&\partial_\eta \ln (Y^3/X)=\partial_\eta \ln a^2\sqrt{V^3/U},\nonumber
\\&=&2\frac{a'}{a}+2\alpha\frac{\exp(\alpha\Delta\eta)}{1-\exp(\alpha\Delta\eta)}, \nonumber
\end{eqnarray}
The evolution equation for the tensor mode is
\begin{eqnarray}
h_{ij}''+2\alpha\frac{\exp(\alpha\Delta\eta)}{[1-\exp(\alpha\Delta\eta)]}h_{ij}'+\frac{\exp(\alpha\Delta\eta)}{[1-\exp(\alpha\Delta\eta)]}k^2h_{ij}=0, \nonumber \\ \label{kgwave}
\end{eqnarray}
A crucial difference with regards to tensor modes in Einstein gravity is that the pre factor of last term in Eq.(12) becomes singular. While in Einstein gravity that term is responsible for the acoustic,
or wave-like behaviour of the tensor mode evolution, in Eddington gravity the Laplacian term is greatly
suppressed. One possible interpretation is that the speed of the gravitational wave goes to $0$ as
$t\rightarrow-\infty$.
We can extract the asymptotic behaviour of the wave equation in
the limit where $\Delta\eta\rightarrow -\infty$ by discarding the last terms in Eq.(12) to find
\begin{eqnarray}
h_{ij}''\simeq 0, \nonumber
\end{eqnarray}
which we can solve to give $h_{ij}\propto A\eta+B$. As announced, we have found an instability in the Eddington regime, in the asymptotic past. The consequences of such a result will be discussed in the conclusions of this report.

We now turn to the case where $\kappa <0$ where we found a bounce when the scale factor of the Universe reaches its minimum, non-singular value. It was shown in \cite{Scargill2012} that, if one choses a closed, positively curved spatial metric, it is possible to construct an oscillating (or \textit{Phoenix} Universe) which undergoes an indefinite number of cycles. Such a model should, in principle allow us to study the evolution of perturbations through the various cycles and shed light on some of the issues that have been raised in the study of cyclic cosmologies \cite{Brandenberger2009,Piao:2009ku}.

Again, we will work deep in the Eddington regime where the $a$, $X$ and $Y$ can be closely approximated by (assuming Euclidean geometry)
\begin{eqnarray}
a&=&a_B(1+\frac{2}{3|\kappa|}t^2), \nonumber \\
X^2&=&\frac{4}{3}a^2\sqrt{\frac{|\kappa|}{2}}\frac{1}{|t|}, \nonumber \\
Y^2&=&\frac{4}{3}a^2\sqrt{\frac{2}{|\kappa|}}|t|, \nonumber
\end{eqnarray}
and we have assumed that the bounce occurs at $t=0$. In conformal time we find that the scale factor
can be expressed as
\begin{eqnarray}
a=a_B[1+\tan^2(\beta\eta)], \nonumber
\end{eqnarray}
where $\beta=a_B\sqrt{2/(3|\kappa|)}$. We then have
\begin{eqnarray}
X^2&=&a^2\frac{4}{3^{3/2}}\frac{1}{|\tan(\beta\eta)|}, \nonumber \\
Y^2&=&a^2\frac{4}{3^{1/2}}|\tan(\beta\eta)|.
\end{eqnarray}
 We can now take the Taylor expansion around $\eta=0$ to find that
 \begin{eqnarray}
 \frac{X^2}{Y^2}=\frac{U}{V}&\simeq&\frac{1}{3\beta^2\eta^2}, \nonumber \\
 \left(3\frac{Y'}{Y}-\frac{X'}{X}\right)&\simeq& \frac{2}{\eta}, \nonumber
 \end{eqnarray}

 The evolution equation for the tensor mode reduces to:
 \begin{eqnarray}
 h_{ij}''+\frac{2}{\eta}h_{ij}'+\frac{k^2}{3\beta^2\eta^2}h_{ij}=0, \nonumber \\ \label{kgwave}
\end{eqnarray}
which can be solved with $h_{ij}\propto \eta^p$ where
\begin{eqnarray}
p=-\frac{1}{2}\pm \frac{1}{2}\sqrt{1-(4k^2/3\beta^2)}. \nonumber
\end{eqnarray}
As in the case of $\kappa>0$ we find an instability as $a\rightarrow a_B$, this time at the bounce; both
solutions blow up as $\eta^{-1/2}$ rendering such a space time unstable to tensor mode perturbations.


\section{Conclusions}
\label{conclusions}
In this paper we found that, even though the background evolution is resolutely non-singular, the overall
evolution is still singular once one considers tensor perturbations. It is an intriguing result and especially
so in the case of $\kappa<0$ where the tensor mode blows up at a finite time. The singular behaviour is
clearly induced by the evolution of the homogeneous part of the auxiliary metric (i.e. via $X$ and $Y$) and
is a completely novel effect, not present in conventional, Einstein gravity.

A similar instability does arise in the radiation era, in Einstein gravity- as we saw above there is
a decaying mode, proportional to $1/k\eta$ at early times and clearly divergent as $\eta\rightarrow 0$. Yet, the current paradigm does not extend the radiation era all the way back to the Big Bang; there is an intervening period of de-Sitter expansion, known as cosmic Inflation, which itself has a finite duration. One
can imagine invoking a more elaborate theory for Eddington theory in the asymptotic past but the simplest scenario, first proposed in \cite{Banados:2010ix}, does not work in its simplest incarnation. The instability is unavoidable for both $\kappa>0$ and $\kappa<0$.

How general is this behaviour? In \cite{BGRS2009} and \cite{Delsate2012}, an alternative view point for these theories of gravity was proposed, firmly placing them in the context of bigravity. Inevitably, in such theories, one finds more than one tensor mode which may or not be tightly coupled to each other, depending on whether the auxiliary metric has its own kinetic term. An analysis of perturbations in \cite{Banados:2008fj} did not find such an instability in the scalar sector and this might be a hint that it is the particular form of the Eddington inspired Born Infeld theory that gives rise to such behaviour. Another possibility is the fact that we are considering a Palatini formulation of gravity where, as shown, pathologies occur that are absent in purely metric theories \cite{Sotiriou2008}. We intend to look at general formulation of gravity theories \cite{Clifton2012} to pin down the conditions in which such an instability arises.

We may learn some lessons from the cosmological setting which may be applicable in other physical circumstances. In particular, there has been some work on understanding the process of gravitational collapse in Eddington gravity. The focus has been on spherical collapse, as one might expect. Clearly traceless, transverse modes may play an unexpected role and should be included if possible  \cite{Liu:2012rc}. Indeed by allowing for more general perturbations it should also be possible trace the effect of nonlinear evolution
of the tensor modes (coupled to radial modes) to search if the singularity can be stabilised in the non-linear regime.

Finally, there is of course the whole realm of gravitational waves to be explored in more general settings in these theories of modified gravity. Our analysis clearly hints at the possibility that interesting effects might arise in these theories in regions of density and curvature. This is clearly one of the new frontiers of modern gravitational physics which merits further exploration, specially given the developments in the study of inspiralling compact objects, pulsars and gravitational wave detection \cite{Psaltis2009}.


\begin{acknowledgments}
We thank T.Baker and T. Clifton for useful discussions.
This research is supported by STFC, Oxford Martin School, BIPAC and
the Basque Government through Ruth Lazkoz's research project AE-2010-1-31,
Fundaci\'on Pablo Garc\'ia, FUNDEC, M\'exico. MB was partially supported by Fondecyt (Chile) Grants \#1100282 and \# 1090753. 

\end{acknowledgments}



\end{document}